\documentclass[manuscript, nonacm]{acmedited}

\usepackage{caption} 
\usepackage{array}
\usepackage{multirow}
\usepackage{makecell}
\usepackage{colortbl}
\usepackage{textcomp}
\usepackage{paralist} % compactitem lists
\usepackage{cleveref}
\usepackage{graphicx}
\usepackage{url}
\usepackage{upgreek}
\usepackage[textsize=tiny]{todonotes}
\usepackage{times}
\usepackage{microtype}
\usepackage{booktabs} % for formal tables
\usepackage{subcaption}

\usepackage[shortcuts]{extdash}

\AtBeginDocument{%
  \providecommand\BibTeX{{%
    \normalfont B\kern-0.5em{\scshape i\kern-0.25em b}\kern-0.8em\TeX}}}

\begin{document}
% TC:macro \cite [option:text,text]
% TC:macro \citep [option:text,text]
% TC:macro \citet [option:text,text]
% TC:envir table 0 1
% TC:envir table* 0 1
% TC:envir tabular [ignore] word
% TC:envir displaymath 0 word
% TC:envir math 0 word
% TC:envir comment 0 0
\title{Toward Finding and Supporting Struggling Students in a Programming Course with an Early Warning System}


\author{Belinda Schantong}
\affiliation{%
  \institution{Chemnitz University of Technology}
  \country{Germany}}
\email{belinda.schantong@informatik.tu-chemnitz.de}

\author{Dominik Gorgosch}
\affiliation{%
  \institution{Chemnitz University of Technology}
  \country{Germany}}
\email{dominik.gorgosch@informatik.tu-chemnitz.de}

\author{Janet Siegmund}
\affiliation{%
 \institution{Chemnitz University of Technology}
 \country{Germany}}
\email{janet.siegmund@informatik.tu-chemnitz.de}

\renewcommand{\shortauthors}{Schantong, et al.}

%RQs
\newcommand{\RQOne}{RQ1: Can the development of an early mental model predict programming skill by the end of a programming course?}

\newcommand{\RQTwo}{RQ2: Can language skills predict programming skill?}

\newcommand{\RQThree}{RQ3: Can selective and sustained attention predict programming skill?}

\newcommand{\RQFour}{RQ4: Can fluid intelligence predict programming skill?}

\newcommand{\RQFive}{RQ5: What skills are suitable for an early warning system?}

\newcommand{\RQSix}{RQ6: Can syntax drill-and-practice exercises improve programming skill?}

\begin{abstract}
 % A clear and well-documented \LaTeX\ document is presented as an
  %article formatted for publication by ACM in a conference proceedings
  %or journal publication. Based on the ``acmart'' document class, this
  %article presents and explains many of the common variations, as well
  %as many of the formatting elements an author may use in the
  %preparation of the documentation of their work.
\textbf{Background}: Programming skills are advantageous to navigate today's society, so it is important to teach them to students. However, failure rates for programming courses are high, and especially students who fall behind early in introductory programming courses tend to stay behind. 

\noindent\textbf{Objective}: To catch these students as early as possible, we aim to develop an early warning system, so we can offer the students support, for example, in the form of syntax drill-and-practice exercises.

% To this end, we assess different cognitive skills that have been identified as correlating with the ability to acquire programming skills.

% We combined different approaches that try to identify students who might struggle in introductory programming courses into a single model, so that instructors may be able to react as early as possible. To make [this] actionable, we also implemented syntax drill-and-practice exercises into our courses.
  
\noindent\textbf{Method}: To develop the early warning system, we assess different cognitive skills of students of an introductory programming course. On several points in time over the course, students complete tests that measure their ability to develop a mental model of programming, language skills, attention, and fluid intelligence. Then, we evaluated to what extent these skills predict whether students acquire programming skills. Additionally, we assess how syntax drill-and-practice exercises improve how students acquire programming skill. 
%A total of between 33 and 111 students participated in the experiments.

% We compared students’ performance in a series of tests and in the syntax exercises to their performance in the final exam of a two-semester introductory programming course. These tests comprised of tests for the development of early mental programming models, language skills, attention, and fluid intelligence. 

\noindent\textbf{Findings}: Most of the cognitive skills can predict whether students acquire programming skills to a certain degree. Especially the ability to develop an early mental model of programming and language skills appear to be relevant. Fluid intelligence also shows predictive power, but appears to be comparable with the ability to develop a mental model. Furthermore, we found a significant positive effect of the syntax drill-and-practice exercises on the success of a course. 

\noindent\textbf{Implications}: Our first suggestion of an early warning system consists of few, easy-to-apply tests that can be integrated in programming courses or applied even before a course starts. Thus, with the start of a programming course, students who are at high risk of failing can be identified and offered support, for example, in the form of syntax drill-and-practice exercises to help students to develop programming skills.

% \noindent\textbf{Future Work}: Further studies are necessary to generalize the results. These could consist of XXX.  
\end{abstract}

%    <concept>
%        <concept_id>10010405.10010489</concept_id>
%        <concept_desc>Applied computing~Education</concept_desc>
%        <concept_significance>500</concept_significance>
%        </concept>
%  </ccs2012>

% \ccsdesc[500]{Applied computing~Education}



\maketitle

%https://icer2022.acm.org/track/icer-2022-papers#Submission-Instructions

\section{Introduction}

% Programming as 4th literacy
% However, many struggle with programming and fail
% Can we help students overcome this struggle?
% We want Frühwarnsystem (early warning system), so that we can offer specific help
% Because we know, that people who fall behind, often stay behind

% We want to know: Können wir vorhersagen, ob jemand programmieren lernt?/
% Vorhersage: kognitiven Fähigkeiten; programmierfähigkeiten
% Können Syntaxaufgaben helfen?\\
% ---\\

% \begin{itemize}
%     \item Bimodal Outcomes
%     \begin{itemize}
%         \item CS1: mehr sehr gute Studierende, aber auch mehr die den Kurs nicht bestehen, als in anderen Kursen \cite{robins_novice_2019}
%         \item Woran liegt das? Robins: eventuell mathematische Skills oder hoher IQ \cite{robins_novice_2019}, bisher aber wenig Untersuchungen
%     \end{itemize}
%     \item Learning Edge Momentum 
%     \begin{itemize}
%         \item Erfolgreiches Lernen erleichtert den Erwerb verwandter Konzepte und macht erfolgloses Lernen schwieriger, wenn ein bestimmter Zielbereich zu erlernender Konzepte vorausgesetzt wird \cite{robins_novice_2019, robins_learning_2010}
%     \end{itemize}
%     \item "One wonders, for example, about teaching sophisticated material to CS1 students when study after study has shown that they do not understand basic loops; more time spent on looping problems might pay a much larger return in the long run." \cite[S. 21]{winslow_programming_1996}
%     \item Eventuell Bezugnahme auf eines der Frameworks, welche Fähigkeiten und "Wissensarten" zum Programmieren benötigt werden?
% \end{itemize}

The ability to program is becoming more and more important every day in a world that is becoming more and more digitalized. Some even see programming as the 4th literacy, as abilities that are so vital to the understanding of our current and future society that everyone will need to acquire them~\cite{roman2017cognitive, vee2013understanding}. Naturally, there is much research on programming education practices, yet we still struggle with teaching students programming skills. The failure rate for introductory programming courses varies around 30\,\% and has not changed much since many years~\cite{luxton-reilly_learning_2016,bennedsen_failure_2007}, despite decades of dedicated research. To address this issues, one line of research studies how cognitive skills are related to programming skills, so that (programming) training can be tailored to how students think and develop programming skills. 
% Researchers sometimes state that learning to program requires a certain aptitude~\cite{bennedsen_failure_2007, robins_learning_2003}, even though evidence for such an aptitude is ambiguous~\cite{jenkins_difficulty_2002, medeiros_systematic_2019}.

% In addition to programming tests, other studies have evaluated predictors of student performance in programming courses by examining cognitive skills that seem to be related to programming skills.
% Several aptitudes have been the focus of studies.
For example, Ambrosio and others found that general intelligence and spatial reasoning are highly related to students' overall performance in a programming course~\cite{ambrosio_identifying_2011}. Additionally, attention to detail also seems to be somewhat related to students' performance, but not mathematical reasoning. Endres and others found that both, reading skill and spatial skill, correlate with programming skill after an 11-week programming course~\cite{endres2021read}.

Rom\'{a}n-Gonz\'{a}les and others used an integrated test battery to assess several cognitive skills, but instead of programming skills, they evaluated how these skills correlate with computational thinking. They also did not recruit university students, but students of middle and high school~\cite{roman2017cognitive}. They found medium (reasoning and spatial skills) to high (problem-solving) correlations, concluding that "computational thinking could be fundamentally linked with general mental ability (particularly with fluid intelligence); and to a lesser extent with different cognitive aptitudes, such as logical reasoning and spatial ability."~\cite[p. 10]{roman2017cognitive}. This could be replicated by Stewart and others with a younger population (i.e., 4th and 5th grade elementary school students)~\cite{stewart2021exploring}. In the end, Rom\'{a}n-Gonz\'{a}les and others have a successfully developed a test to measure computational thinking on the level of 5th to 10th grade students and are able to identify computationally talented students already in middle school~\cite{roman2018predictValidity}. However, struggling students cannot be identified with this test. 

% Rom\'{a}n-Gonz\'{a}lez and others also used standardized psychological tests in the process of validating their test to measure computational thinking~\cite{roman2017cognitive}. They found high correlation between their test and the Primary Mental Abilities battery, which consists of four different subtests that measure verbal, spacial, reasoning and numerical abilities~\cite{pma2007}. Of these four, spacial and reasoning ability showed the highest relationship with computational thinking. They also correlated the results of their computational-thinking test with a problem-solving test that is highly associated with general mental ability. They conclude that "computational thinking could be fundamentally linked with general mental ability (particularly with fluid intelligence); and to a lesser extent with different cognitive aptitudes, such as logical reasoning and spacial ability."~\cite[p. 10]{roman2017cognitive}

Helmlinger and others also evaluated reasoning skills on different modalities (verbal, numeric, figural), and found that only figural reasoning had a relationship with programming experience, such that students with programming experience performed significantly better in figural inductive reasoning tasks than students with no programming experience~\cite{helmlinger2020programming}.

Prat and others assessed an entire battery of cognitive skills and computed a stepwise regression model of how these skills can predict programming skills~\cite{prat2020relating}. They found that numerical skills seem to have no influence on acquired programming skills, but, depending on the concrete operationalization of programming skills, either general cognitive abilities (specifically fluid reasoning, working memory updating and working memory span) or language aptitude seem to be relevant~\cite{prat2020relating}.

% Additionally, they also found that programmers' allocation of neural processing during the tasks was more efficient.
These results are in line with recent neuro-imaging studies, in which a considerable contribution of language-processing areas in the brain are observed~\cite{siegmund_understanding_2014, siegmund_measuring_2017, floyd_decoding_2017, lee_comparing_2016}. Additionally, the contribution of spatial abilities has also been shown~\cite{huang2019distilling}. Thus, there is strong evidence of connection of several cognitive skills to programming.

% For learning rate (i.e., how many lessons students managed to complete within 10 training sessions), language aptitude is a relevant predictor, which is in line with several neuro-imaging studies showing a clear involvement of language-specific brain areas when students work with code~\cite{siegmund_understanding_2014, siegmund_measuring_2017, floyd_decoding_2017, lee_comparing_2016}.

% Luxton-Reilly et al.\~\cite{luxton-reilly_learning_2016}(2016) reported that the failure rate for introductory programming courses in New Zealand universities is 15\% higher than for other introductory courses. This fits the numbers of Bennedsen and Caspersen\cite{bennedsen_failure_2007} who report a failure rate of 33\% in introductory programming courses worldwide. Education in computer science has been failing to address these problems for 40 years, claiming that programming is just difficult to learn or that there simply is a certain aptitude needed to learn it, even though evidence for such an aptitude is ambiguous~\cite{jenkins_difficulty_2002}.

% In terms of the neo-Piagetian model, which consists of a sensorimotor - the student's model of program execution is wrong, Preoperational - students can trace single lines of code but do not understand the purpose of an entire program, and Concrete operational - students can read and understand code and are able to write simple algorithms for the first time~\cite{teague_longitudinal_2014, lister_concrete_2011}
Not only cognitive skills have been used to assess programming skills or to predict whether someone will acquire skills, also early programming tests have been developed and applied. For example, Ahadi and others evaluated how the development of a suitable mental model\footnote{In terms of the neo-Piagetian model, which consists of a sensorimotor, preoperational, and concrete operational stage~\cite{lister_concrete_2011}} for basic programming concepts (i.e., how values are assigned to and moved between variables) is related to the success or failure of a programming course~\cite{ahadi_falling_2014}. They found that students who do not develop a suitable mental model at the beginning of a programming course often cannot overcome these difficulties.

% In summary, many students are struggling with learning to program, and the question is how we can help students overcome this struggle. In a series of studies, we have been aiming to build an ‘early warning system’. Our goal is to find struggling students as early as possible, to be able to take according measures, since evidence shows that people who fall behind early tend to stay behind for good~\cite{ahadi_falling_2014}.

% mathematical reasoning and attention to detail and are related to students' performance in a programming course.They found large correlation effects between the results and the tests for general intelligence and spatial reasoning. The test for the attention to detail showed a medium sized correlation effect and the test for mathematical reasoning only a small one. \todo[inline]{Belinda: Add the other papers with the tests here, similar to Ambrosio. You can also start with bullet points}

% Prat et al. examined participants in a Python course for beginners. They correlated the course outcomes with the results from tests for language aptitude, numeracy and general cognitive abilities (specifically fluid reasoning, working memory updating and working memory span). They found that language aptitude and general cognitive measures on average explained around half of the variance in python learning outcomes, though language aptitude was more relevant in explaining the participants' learning rate, while cognitive ability was more relevant in explaining the participants' accuracy in programming. Numeracy on the other hand had very little predictive utility.

Inspired by this line of research, we set out to develop an early warning system for students who are at high risk of failing a programming course. Since learning to program has its difficulties, it is important to intervene early enough so students do not fall behind early and stay behind, but instead can close their potential mental gap (or even avoid its emergence). To this end, we assess several of the cognitive skills and early programming skills that have been subject of previous studies.
% Specifically, we reuse the programming tasks from Ahadi and others, as they appear reliable to predict programming skills~\cite{ahadi_falling_2014}. Additionally, we collect performance in several cognitive skills, including language skills and fluid intelligence,
Our goal is to combine the results in a statistical model that predicts the programming performance of students at the end of a programming course.

With our early warning system, we would have an instrument that would help us to identify struggling students at a point where we would still be able to implement counter measures, such as tailored assignments~\cite{thurner2017concept} or training of skills that are related to programming and can be transferred to programming skills~\cite{endres2021read}.

Another approach are syntax drill-and-practice exercises,
% to help students use their cognitive resources more efficiently and free them from struggling syntax learning, which.
% to other important parts of learning programming.
% To this end, we aim to free cognitive resources that are bound by learning the syntax of a language, by integrating syntax drill-and-practice exercises into our courses.
since novice programmers often have problems with learning the syntax of a programming language, a phenomenon known since the 80s~\cite{soloway_cognitive_1983}. These problems persist to this day, even for students who otherwise show good performances in programming. For example, Denny and others found that nearly half of all submissions by students in the top quartile contain syntax errors~\cite{denny_understanding_2011}. Based on these findings, several studies used syntax drill-and-practice exercises to reduce the cognitive load caused by uncertainty about the syntactical rules of a programming language~\cite{edwards_separation_2018, edwards_syntax_2020, ly_revisiting_2021, sullivan_student_2021}. We integrated syntax drill-and-practice exercises into the second part of a programming course to evaluate whether they can improve programming skills of students at the end of the course.

In a nutshell, we found that especially the development of a suitable early mental model and language skills are potentially useful predictors for our early warning system. Fluid intelligence might also have some predictive power, but not attention. In summary, we make the following contributions:

% Concerning syntax exercises, we find that students who participated in the drill-and-practice exercises performed significantly better in the final exam of the course.

% In addition to our findings described in this paper, we want to provide our materials and results online to facilitate replication.

\begin{itemize}
    \item Empirical evidence for the role of cognitive skills in learning to program, strengthening the result of previous studies
    \item Differentiation of results of previous studies, showing ceiling effects in the cognitive skills of students of a programming course
    \item Course-specific syntax drill-and-practice exercises that improve programming skills
    \item The experimental design and raw data is available (\url{https://anonymous.4open.science/r/9252151/}) for others to build on and replicate our results, and we explicitly encourage researchers of other programming courses to do so.
\end{itemize}

\section{Experiment} \label{experiment}
\subsection{Objective Definitions}

To guide our study design, we formulated 6 research questions. For each of these questions, programming skill is the dependent variable, which is operationalized as the performance (in terms of score) in the final exam of the programming course. In Section~\ref{sec:material}, we provide more details on this exam, as well as all independent variables.

%study consists of several experiments - EIOS tasks, language and cognitive tests, and syntax tasks. The goal of this study is to create an early warning system that allows us to identify students who might fall behind in introductory programming courses and to investigate whether a simple to implement intervention like syntax tasks can lead to better results in programming learning. Specifically, we pose the following research questions:

\textbf{\RQOne}

Ahadi and others have shown that a set of eight simple programming tasks can predict whether students without programming experience successfully complete a programming course~\cite{ahadi_falling_2014}. These programming tasks, being provided a few weeks into the course, evaluated whether students have learned the concept of variables as having a state, rather than being a box that contains information. The easiest task was to identify the value of a variable after a few assignment statements, the most difficult task asked students to implement a swap of values between two variables. The cognitive prerequisites to successfully complete these tasks can be mapped to the neo-Piagetian stages of learning, with the easiest task mapping to the sensorimotor stage, and the most difficult to the concrete operational stage~\cite{lister_concrete_2011}. To investigate whether these tasks could be integrated into our early warning system, we replicated the study by Ahadi and others, comparing the students' performance in these tasks at the beginning of the course to their performance in the final exam.

% applying neo-Piagetian theory to programming learning resulted in reliable predictors of programming learning. They created a test consisting of assignment statements, one of the earliest semantic models needed in programming. They were able to assign students to different stages of the neo-Piagetian model based on their performance in those tasks. To investigate whether these tasks could be integrated into our early warning system, we replicated the study by Ahadi and others, comparing the students performance in their tasks at the beginning of CS1 to their performance in the final exam of CS2.

\textbf{\RQTwo}

\textbf{\RQThree}

\textbf{\RQFour}

Since all these cognitive skills have been identified in some studies as predictor for learning to program, we include these in our study. The advantage of using purely cognitive skills that do not require knowledge of programming is that we can assess them even before a programming course starts, so that we might be able to identify possibly struggling students even before they struggle.

\textbf{\RQFive}

Previous studies have shown varying levels of predictive power for all the skills mentioned in RQ1 to RQ4. Some correlations are promising (e.g., between the Computational Thinking test and problem-solving ability with r = 0.67~\cite{roman2017cognitive}, or between language aptitude and programming learning rate with r = 0.56~\cite{prat2020relating}). However, this means that only some portion of the variance in the acquired programming skill can be accounted for. To increase the predictive power of these skills, we aim at combining them (i.e., the development of a suitable mental model (RQ1), language skills (RQ2), selected and sustained attention (RQ3), and fluid intelligence(RQ4)) into one model.

\textbf{\RQSix}

To provide an actionable plan for educators, we integrate syntax drill-and-practice exercises to our programming courses. Previous studies have shown the success of such exercises in improving the learning outcome of a programming course~\cite{edwards_separation_2018, edwards_syntax_2020, ly_revisiting_2021, sullivan_student_2021}. Given the consistent finding of language skills as predictor for programming skill, and that learning syntax is difficult~\cite{denny_understanding_2011} and coupled to language skills, this seems to be a promising option as intervention for students who appear to have fallen behind in a programming course.

\subsection{Material}
\label{sec:material}

In this section, we present the operationalization of our variables, arranged by research question. We start with the dependent variable, the programming skill.

\subsubsection{Programming Skill}

As indicator for programming skill, we use the points that students receive in the final exam. The format was a take-home exam. Students received instructions to implement three programs. We defined four tasks in total, and students could choose which of Task 3 or 4 to implement.
    \begin{itemize}
        \item Task 1: Implementation of a double linked list for integer values, including an algorithm that successively deletes sub-lists whose sums equal a prime number.
        \item Task 2: Implementation of a tree structure with weighted nodes, including an algorithm that searches for all nodes within a certain range of weight, while also considering the weight of their child-nodes.
        \item[] \emph{and either one of:}
        \item Task 3: Implementation of an algorithm that finds the shortest path on the board of a game while considering different types of terrain and different movement-speed assigned to units in the game.
        \item Task 4: Implementation of the simulation of a social network, including methods to identify friendship-circles within the network.
    \end{itemize}
    % Klausur mit den restlichen Materialien online zugänglich machen

The time limit to complete the tasks was six hours. Students could implement their solutions in either Java or Python; both languages were used in the programming course. Since this was an open-book exam, students were allowed to use materials and code from the course and the internet, but needed to work alone.

While the conditions of the exam are less controlled than in a closed-book exam under observation, the measurement of programming skills is much more realistic. Thus, although there might be some outliers among the students about how they conduct the exam (e.g., seek help of colleagues), we believe that, for the majority of students, we have a valid operationalization of programming skill.
 
\subsubsection{\RQOne}

To evaluate whether and how students have developed an early mental model of programming, we used the Early-mental-model test (Early-MM) created by Ahadi and others~\cite{ahadi_falling_2014}. It consists of eight tasks that test the student's understanding of assignment statements, one of the first concepts required to learn in programming, with the most difficult one requiring to implement a value swap between two variables. For each task, students get either 0 or 1 point, so the minimum of this test would be 0 points and the maximum 8 points.

\begin{figure}
    \centering
    \begin{subfigure}{0.49\textwidth}
      \includegraphics{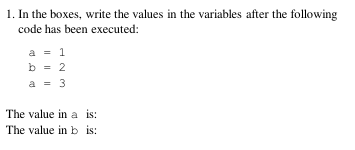}
      \caption{The easiest task was to name the values of variables a and b (a = 3 and b = 2 is correct).}
    \end{subfigure}
    \begin{subfigure}{0.49\textwidth}
      \includegraphics{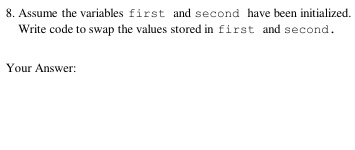}
      \caption{The most difficult task was to implement a swap. A correct solution would be \texttt{temp = first; first = second; second = temp}}
    \end{subfigure}
    \caption{Tasks of the Early-MM~\cite{ahadi_falling_2014}.}
    \label{fig:EIOS}
\end{figure}

In Figure~\ref{fig:EIOS}, we show the easiest and most difficult tasks with correct solutions. Students who can implement the swap correctly must have fully understood the concept of variables having a state, whereas students who fail to answer the easiest question correctly have not yet understood how variables work. However, since variables are the building blocks of most programs, students will struggle with keeping up understanding higher concepts. Thus, these tasks may be a suitable predictor for programming skills.

\subsubsection{\RQTwo}

%anonymous
Since language skills are reliably found to predict programming skill, we selected several language skill tests to have different options to explore a relationship with programming skills. All three tests were provided by a local institute specialized in test development (will be disclosed after reviewing).
%the Institute for Test Research and Test Development Leipzig (\url{https://itt-leipzig.de}).

\begin{figure}
    \centering
    \begin{subfigure}{0.49\textwidth}
      \includegraphics[width=\textwidth]{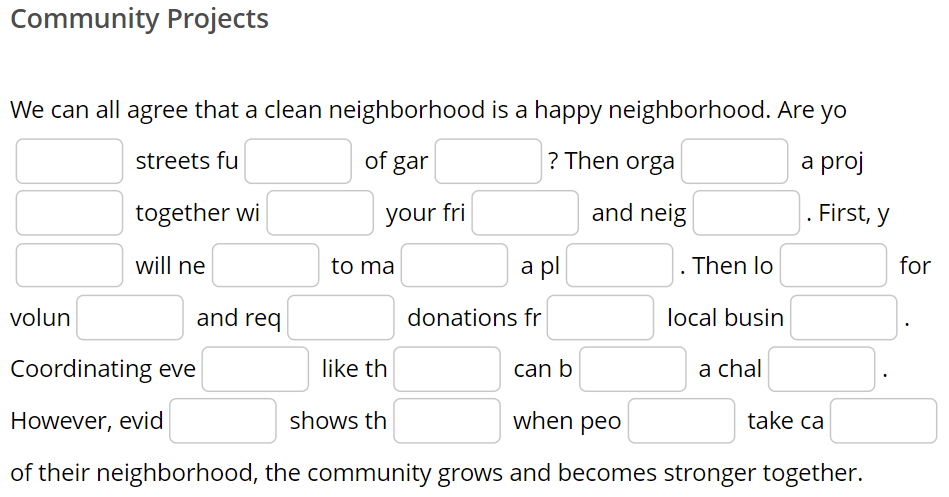}
      \caption{As shown to students.}
    \end{subfigure}
    \begin{subfigure}{0.49\textwidth}
      \includegraphics[width=\textwidth]{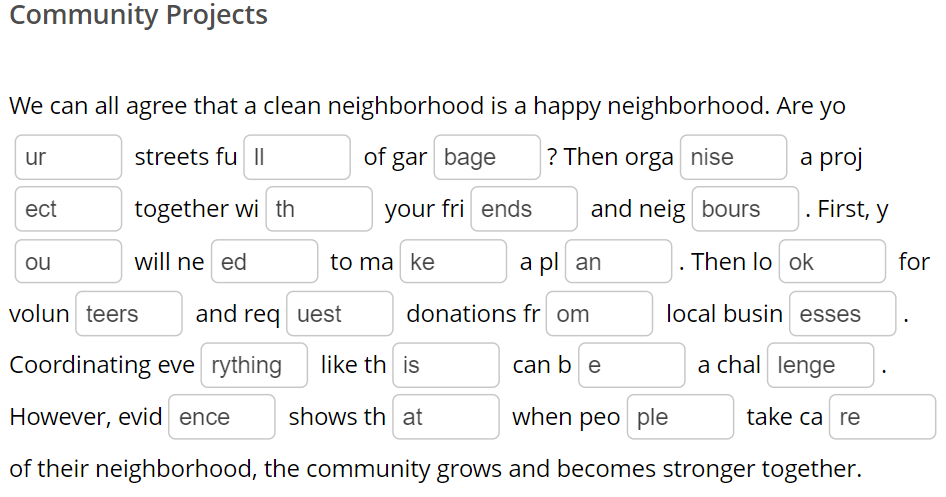}
      \caption{Version with solutions (not shown to students).}
    \end{subfigure}
    \caption{One item in the C-Test.}
    \label{fig:c-test}
\end{figure}

\paragraph{C-Test (English)}
%anonymous
A C-Test is an integrative written test to measure overall fluency in a specified language and is usually used to assess the participant's reference level in the Common European Framework of Reference for Languages (CEFR)\footnote{The CEFR describes foreign language proficiency at three divisions, each divided into two levels: Basic users (levels A1-A2), independent users (levels B1-B2) and proficient users (levels C1-C2), with the highest level (C2) being similar to a native speaker. For further explanation, see~\cite{cefr}.}~\cite{ajideh2012ctest}. The C-Test specifically ranked participants from the CEFR levels B1 to C1. For comparison, the level expected from high-school graduates is B2, with high-performing students being able to reach C1~\cite{keller2020english}. The C-Test has four items which consist of a cloze text in the target language that is constructed according to specific rules. Each text has 25 gaps. We show an example with solution in Figure~\ref{fig:c-test}. The participant's score is computed by counting the correctly solved gaps, leading to a maximum score of 100 points. Participants had 5 minutes to solve each item.
% The specific test used here placed participants in the CEFR levels from B1 to C1. 

\paragraph{Receptive and Productive Vocabulary Test (English)} 

\begin{figure}
    \centering
    \begin{subfigure}{0.49\textwidth}
      \includegraphics[width=\textwidth]{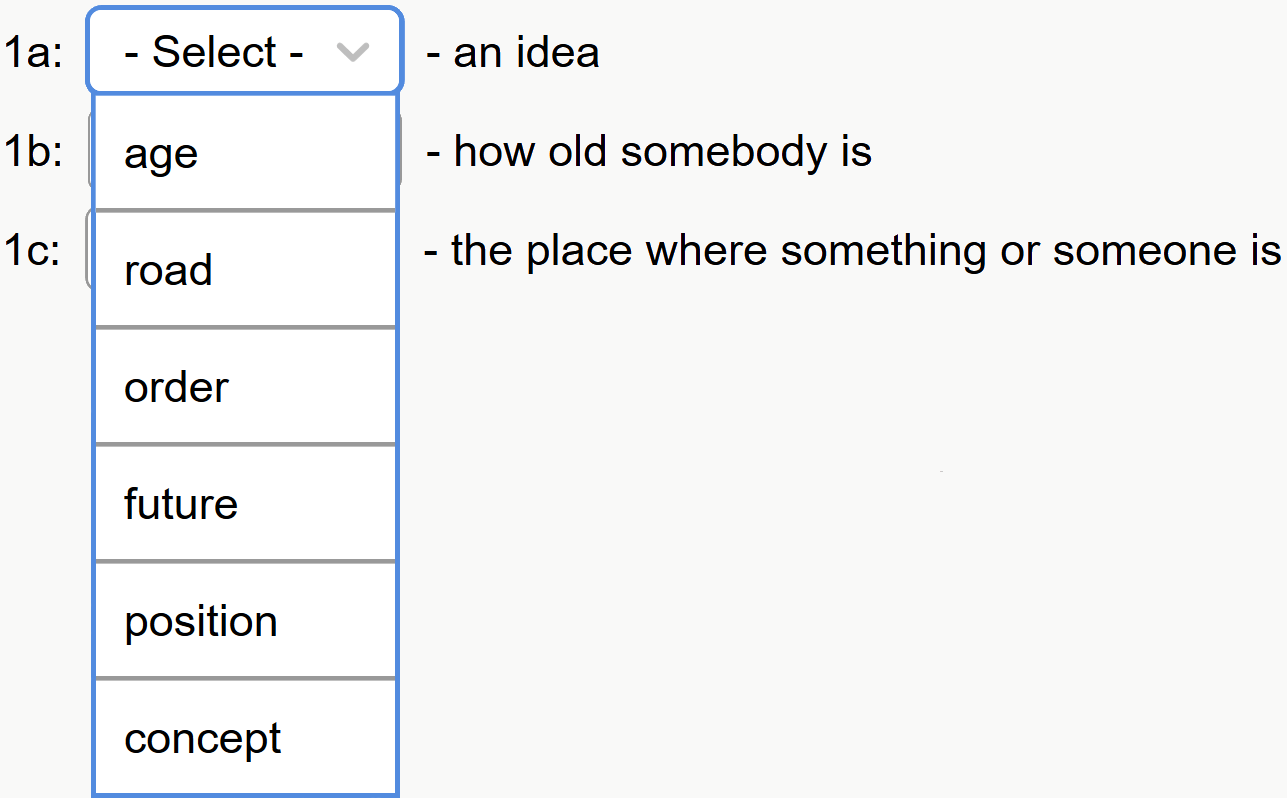}
      \caption{For question 1a, 'concept' needs to be selected from the drop-down menu.~\cite{wortschatz_itt}}
      \label{fig:vocs-rec}
    \end{subfigure}
    \begin{subfigure}{0.49\textwidth}
      \includegraphics[width=\textwidth]{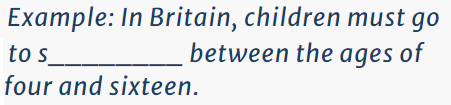}
      \caption{The solution is 'school', thus 'chool' needs to be added.~\cite{wortschatz_itt}}
      \label{fig:vocs-prod}
    \end{subfigure}
    \caption{Possible items of the R-Voc (a) and the P-Voc (b).}
    \label{fig:vocs}
\end{figure}

%anonymous (our country vs. Germany)
The vocabulary tests assess vocabulary size and indicate the test person’s reading level in the target language. They assess the participant’s vocabulary knowledge in five different ranks, which are based on high-frequency vocabulary lists. The first rank assesses knowledge among the 1000 most frequent words in the target language, the second rank among the most frequent 2000, and so forth, up to rank five~\cite{wortschatz_itt}. A participant is considered to have passed a rank if they reach at least 80\,\% of the possible points in the rank. From the passed vocabulary ranks, the person's reading level in the CEFR can be inferred, in this case, ranging from level A2 to B2, so narrowly reaching the level expected from high-school graduates~\cite{keller2020english}. We used both, a receptive and a productive vocabulary test. Receptive (or passive) vocabulary competence refers to the words a person is able to understand while reading, which is usually higher than the productive vocabulary competence, which refers to words that can be actively used~\cite{wortschatz_itt}. Points are awarded for both tests by counting the correctly solved tasks. Participants had a time limit of 30 minutes per test. Both tests use similar, but different items:
\begin{itemize}
    \item An item of the receptive vocabulary test (R-Voc) consists of a drop-down menu with six words from the current vocabulary level, and three phrases that define one of the words from the drop-down menu, as shown in Figure~\ref{fig:vocs-rec}. The test person matches the words with the corresponding phrases. The tests consisted of 10 items per rank with up to 3 points per item, leading to maximum 30 points per rank and a total possible score of 150. 
    \item An item of the productive vocabulary test (P-Voc) consists of a cloze text that is one or two sentences long. There is one gap for the target word, as shown in Figure~\ref{fig:vocs-prod}. To disambiguate an item, up to the first three letters of the missing word are given. The test consists of 18 items per rank with 1 point per correctly solved item, leading to a maximum of 18 points per rank and a total possible score of 90.
\end{itemize}

% \begin{figure}
%     \centering
%     \includegraphics[width=.5\columnwidth]{figures/p_voc_example.png}
%     \caption{Example of a possible item in the P-Voc. The solution is 'school', thus 'chool' needs to be added.~\cite{wortschatz_itt}}
%     \label{fig:p_voc}
% \end{figure}

% \begin{figure}
% \centering
% \begin{subfigure}{0.49\textwidth}
%         \includegraphics[width=\textwidth]{figures/barplot_score.pdf}
%           \caption{}
%           \label{fig:barplot_score}
%       \end{subfigure}
%       \begin{subfigure}{0.49\textwidth}
%         \includegraphics[width=\textwidth]{figures/barplot_per_task.pdf}
%           \caption{}
%           \label{fig:barplot_per_task}
%       \end{subfigure}
% \caption{Barplot - \ref{fig:barplot_per_task} Percentage of correct solutions per task, Percentage of participants per sum of points}
% \label{fig:histogram}
% \end{figure}

\subsubsection{\RQThree} 
\begin{figure}
    \centering
    \includegraphics[width=.7\columnwidth]{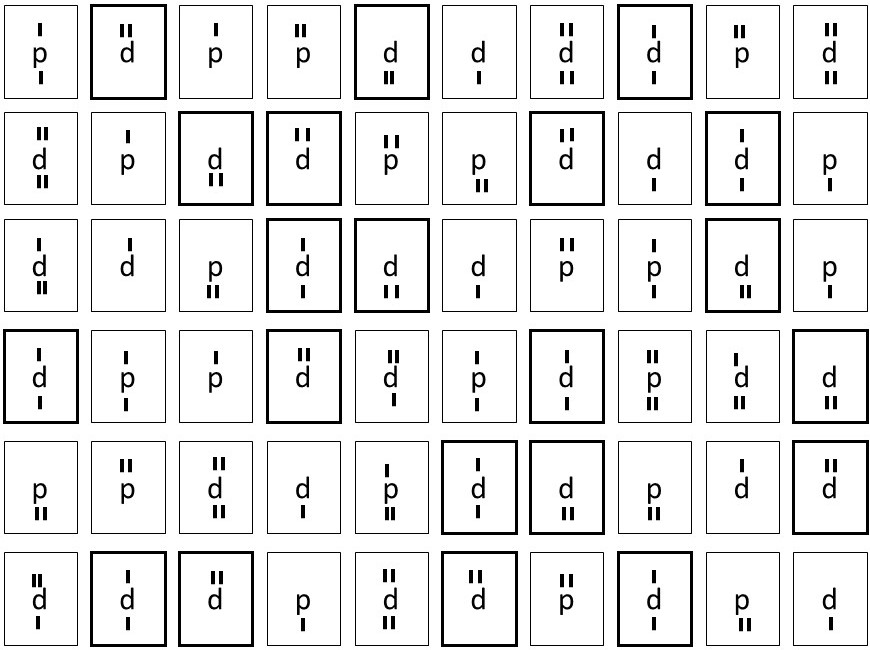}
    \caption{The task is to react as quickly as possible to any d with two lines. The target symbols are highlighted with thicker boxes for purpose of illustration.}
    \label{fig:d2}
\end{figure}

The d2 Test of Attention measures selective and sustained attention, and visual scanning speed, allowing for a neuropsychological estimation of individual attention and concentration performance~\cite{brickenkamp2016d2}. For this study, we used the digital version of the d2-R (Revision). In Figure~\ref{fig:d2}, we show an excerpt of the test. Participants were instructed to react as quickly as possible to every symbol containing a d with two lines, no matter where the lines were. The test consists of 14 items, with a time limit of 20 seconds per item. One item consists of 6 rows, each row consisting of 10 symbols. Including instructions, the test can take up to 10 minutes to complete, but typically takes less time. The results of the d2-R are measured on a T scale, so the standardized mean is 50, with a standard deviation of 10. The test's reference groups consists of Europeans, male and female, aged 18 to 55~\cite{brickenkamp2016d2}.

\subsubsection{\RQFour} 
\begin{figure}
    \centering
    \includegraphics[width=.3\columnwidth]{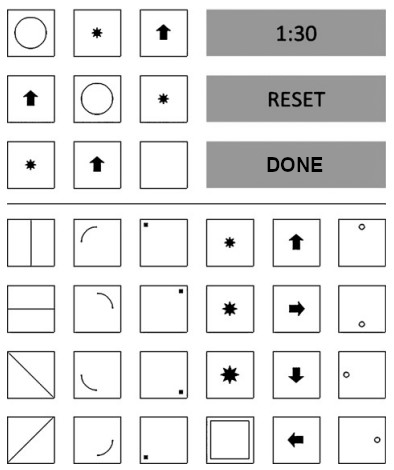}
    \caption{DESIGMA-A: Participants need to compose the missing figure of the matrix. In this example, the correct solution is a circle, which can be composed by the elements of the second column.~\cite{becker2014matrizenkonstruktionsaufgabe}}
    \label{fig:desigma}
\end{figure}

The DESIGMA-Advanced is a test for fluid intelligence using figural matrices. In Figure~\ref{fig:desigma}, we show one item of the test. The missing figure is a circle, which participants can compose with the elements of the second column. With this composition approach, distractor elements are not necessary, so the probability that a participant can guess the solution by process of elimination is reduced~\cite{becker2014matrizenkonstruktionsaufgabe}.

There are 38 items in the entire test, and the time limit is 90 seconds per item, meaning the test takes at max 57 minutes to complete. The results of the DESIGMA-A are measured on the IQ scale, so it has a standardized mean of 100 and a standard deviation of 15.

\subsubsection{\RQSix}
%When learning a new natural language or even a programming language, it is common for learners to make syntactic errors \cite{qian_students_2017}.

\begin{figure}
\captionsetup[subfigure]{justification=centering}
    \centering
      \begin{subfigure}{0.49\textwidth}
        \includegraphics[width=\textwidth]{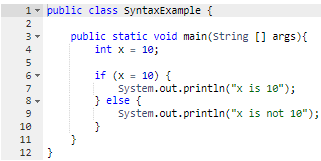}
          \caption{}
          \label{fig:if_exercise}
      \end{subfigure}
      \begin{subfigure}{0.49\textwidth}
        \includegraphics[width=\textwidth]{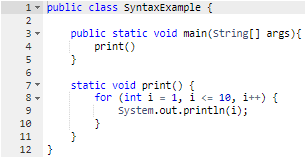}
          \caption{}
          \label{fig:for_exercise}
      \end{subfigure}
      \begin{subfigure}{0.49\textwidth}
        \includegraphics[width=\textwidth]{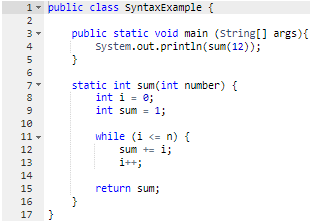}
          \caption{}
          \label{fig:while_exercise}
      \end{subfigure}
      \begin{subfigure}{0.49\textwidth}
        \includegraphics[width=\textwidth]{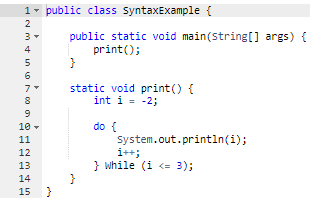}
          \caption{}
          \label{fig:do_while_exercise}
      \end{subfigure}
      \caption{Syntax drill-and-practice exercises. The errors are: (a) Line 6 - Incorrect Assignment Operator (Category: Other), (b) Line 8 - Incorrect separators in the head of a for-loop (Category: Semicolon), (c) Line 14 - Missing closing bracket (Category: Bracket), (d) Line 13- incorrect capitalization of a keyword (Category: Capitalization).}
      \label{fig:syntax_exercises_example}
\end{figure}

We designed syntax drill-and-practice exercises for the following programming constructs: conditions, for-, while- and do-while-loops. We introduced syntax errors in each of the constructs. We defined 5 categories of errors: brackets (mismatched, unbalanced, or abused usage of brackets, braces and parenthesis), semicolon (e.g., an unnecessary semicolon, or an incorrect character instead of a semicolon), loop counter (e.g., i+ instead of i++), keywords (e.g., capitalized keywords) and several other errors that do not fit into the categories (e.g., a misplaced else if). In Figure~\ref{fig:syntax_exercises_example}, we show examples for different constructs and different error categories.

The tasks were summarized per programming construct, leading to 23 to 25 tasks per construct (98 tasks in total), with varying categories of errors. For each construct, we expected students to take about 15 minutes to complete, but there was no time limit for the exercises. Students should identify and fix errors so that a program would compile; each task contained one error. The responses were assessed automatically, giving the participants direct feedback. Students could obtain either 0 or 1 point for a task, leading to a maximum of 98 points. We presented the programming constructs in fixed order, but the individual tasks per construct were randomized.

%\begin{figure}[!ht]
%  \centering
%  \subfloat{\includegraphics[width=0.45\textwidth]%{figures/Bild1.png}a}
%  \subfloat{\includegraphics[width=0.45\textwidth]%{figures/Bild1.png}}
%  \subfloat{\includegraphics[width=0.45\textwidth]%{figures/Bild1.png}}
%  \subfloat{\includegraphics[width=0.45\textwidth]%{figures/Bild1.png}}
%\caption{Test}
%\label{fig:syntax_exercises_examples}
%\end{figure}

%[Bild3\label{subfig-3:Bild3}]

%\begin{figure}
%    \centering
%    \includegraphics[width=0.8\textwidth]{figures/%SyntaxExerciseExample.png}
%    \caption{Syntax exercises example. The errors %are in Line 9, 10 and 11.}
%    \label{fig:syntax_exercises_example}
%\end{figure}

% In Figure~\ref{fig:syntax_exercises_example}, we show an example. The tasks contained syntax errors that the students had to fix so that the program could be compiled. Types of errors in the syntax drill-and-practice exercises were for example: Missing a parenthesis, incorrect capitalization of a keyword, missing semicolons, missing or wrong separators in the head of a for-loop, or a missing counter variable in a while- or do-while-loop.

\subsection{Participants}

As participants, we recruit students from a programming course that spans two semesters. The first part of the course is referred to as \textit{Algorithms and Programming} (AP) and the second part is called \textit{Data Structures} (DS). Due to the pandemic and varying study regulations, we have a high fluctuation of participants for the different research questions. Specifically, 111 students participated for RQ1, and for the remaining RQs, we have between 35 and 50 participants. The age ranged between 18 to about 43. Students who started AP have been programming for an average of about 1.8 years (sd = 1.8) before taking the course. Most students are enrolled in a major that is rooted in the department of computer science at our university, with some of the majors being more interdisciplinary. All students are non-native English speakers. A detailed breakdown of students based on each research question is available on the project's Web site.

\subsection{Conduct}
After consulting with the instructors of AP, the Early-MM test was released to the students shortly after the start of the first semester. It was given in the form of an online survey, which was available to the students over the course of two months. As compensation, students received 2 bonus points for their assignments in AP for participation (independent of the performance).

% The examination of CS1 consisted of seven programming tasks that were given over the course of the semester. Students had two weeks time to work on each task and could achieve a total of 52 points. Students had to get at least 26 points to pass the course. Alternatively, it was also possible to pass the course by taking an exam at the end of the semester. However, we were not able to get a hold of the exam results.

In DS, the experiment was linked with assignments that were mandatory to pass for the students to be allowed to take part in the exam. Students would need to get at least 50\,\% of the points that were achievable in the assignments to pass. Students could choose whether they wanted to participate in the experiment or not. If they chose to participate, they would get slightly easier assignments to make up for the extra time they would use for the experiment, but they still had access to the normal version of the assignments in case they needed it, for example, in preparation for the exam. In addition, students who participated in all tests were offered two bonus points to count to the 50\,\% of points, independent of their performance in the tests. Few students exploited this by just clicking through tests, which we removed from our analysis (we present details in the results section).

The course spanned fourteen weeks. Assignments were distributed weekly starting in the third week of the course. The students had two weeks to complete each assignment, including the tasks for the experiment. The tests and syntax drill-and-practice exercises were distributed as follows:
\begin{itemize}
    \item Week 7: Syntax drill-and-practice exercises
    \item Week 9: d2-R and C-Test
    \item Week 10: DESIGMA-A, R-Voc and P-Voc
\end{itemize}

One could argue that applying the cognitive tests to be used as early warning systems so late in the course does not make much sense. However, since in this experiment, we are developing the early warning system (not applying it), and since the cognitive skills as measured by these tests are not subject to change considerably over the course of several months, the results would still tell us whether these tests are useful for an early warning system. In that case, they must be applied early on, possibly even before the start of a course.

The final exam took place five weeks after the end of the fourteen-week course. 

\section{Threats to Validity}
\label{threats_to_validity}

\subsection{Construct Validity}
With measuring cognitive constructs, it is always challenging to find a suitable operationalization that is also fast enough to apply. With this compromise, we have used for most cognitive skills established tests that are fast to apply; however, we found that they cannot differentiate well among the high-skilled population of students. Thus, we might misinterpret the role of certain cognitive skills as predictor for programming skill. To mitigate this threat, we take it into account when discussing the results.

For programming skills, one could also argue whether the final exam is a good operationalization. However, since the exam takes place in a realistic setting with sufficient time, we believe that our measurement of programming skill is valid (e.g., evaluation apprehension is reduced).

\subsection{Internal}

Since participation was voluntary, we have a high risk of selection bias. Given the number of tests, distributed over two semesters, we have only a relatively small number of students who participated in all tests and the final exam. Thus, our results for the early warning system should be considered in light of this higher motivation.

Additionally, students did all the tests at home, online, and without supervision. Thus, we cannot be sure that all students followed the protocol for all tests. To mitigate this threat, we removed outliers that indicate when students violated the instructions (e.g., were considerably faster than the expected time a test takes).

\subsection{External Validity}

It is difficult to generalize the results to other settings, such as different programming courses, different majors, or different education systems. Nevertheless, since our sample is representative for many universities in our country, they have an implication for many students who want to learn programming.

\subsection{Statistical Conclusion Validity}

Unfortunately, due to the nature of the long-term experiment, and also the Covid-19 pandemic, our  sample on which we intended to compute the integrated model with the cognitive skills is too small; only 18 student completed all the tests. For the other tests, the data is rather fragmented, so we also could not select a useful subset of tests for cognitive skills. Thus, we cannot reasonably compute a statistical model. Instead, we only did a manual analysis of the correlations and co-correlations of our variables to not come to false conclusions.

\section{Results} \label{results}

\Crefname{figure}{Fig.}{Figs.}

% helpers

Since we have highly fragmented data, (i.e., not all students completed all cognitive tests and the exam), we decided to present the data description per research question in 2 steps. For completeness, we first present the data of all students who completed a test that is used for a certain RQ. Second, we present the data for the students who participated in an RQ \textbf{and} who completed the final exam. For example, for RQ1, 111 students took the Early-MM test, but only 32 of these took the final exam. For outlier removal, we report the concrete procedure for each of the research questions separately, since there was no unified procedure due to the diverse nature of the tests.

% \subsection{Preprocessing}
% Ganzheitlich schwierig, weil wir es nicht "mandatory" vorgeben können 

% \begin{itemize}
%     \item Syntaxaufgaben
%     \begin{itemize}
%         \item Aufteilung in Teilnehmende und den Rest zur Auswertung der Klausurergebnisse
%     \end{itemize}
% \end{itemize}

% \subsection{Descriptive Statistics}

Before diving into the research questions, we start by presenting data on the final exam. A maximum of 100 points could be reached, and it was taken by 102 students. The mean was 66.61 with a standard deviation (sd) of 18.16. The median was 68.5 points. Since the number of students who took part in the tests for the RQs fluctuated, we present the concrete mean of the participants in the exam per research question in Table~\ref{tab:summary}. The exam results were not normally distributed, as assessed by the Shapiro-Wilk-Test (p = 0.004). Thus, we always the Spearman rank correlation as non-parametric alternative to the Pearson correlation when comparing test results. 

% Out of these 102 students, 39 participated in the study. Of these 39, the mean was 73.79 (sd = 14.31) and the median was 74 points.

\begin{center}
\begin{table}
    \caption{On the left: Mean results for each test with all participants. On the right: Mean values of the exam and the tests, as well as correlation values between the exam and each test, with participants who did took in both.}
    \label{tab:summary}
    \begin{tabular}{lp{.1\textwidth}r||p{.12\textwidth}rrp{.1\textwidth}}
        \hline
         Test & Participants (All) & Mean Test (SD) & Participants (Test + exam) & Mean Exam (SD) & Mean Test (SD) & Correlation with exam \\
         \hline
        %  Final exam & 102 & 66.61 (18.16) & 39 & 73.79 (14.31) & --\\
         Final exam & 102 & 66.61 (18.16) &  &  &  &\\% 
         \hline
        %  \addlinespace
         Early-MM & 111 & 6.17 (1.86) & 32 & 75.09 (15.22) & 6.97 (1.03) & 0.23\\
         C-Test & 50 & 76.90 (14.62) & 36 & 73.75 (14.79) & 80.11 (12.03) & 0.20\\
         R-Voc & 38 & 129.76 (18.09) & 27 & 73.65 (14.99) & 130.74 (14.41) & 0.28\\
         P-Voc & 35 & 61.77 (11.45) & 26 & 72.81 (14.69) & 62.38 (11.85) & 0.27\\
         d2-R & 49 & 57.12 (8.99) & 36 & 73.75 (14.79) & 57.11 (9.27) & 0.09\\
         DESIGMA-A & 35 & 115.97 (12.50) & 25 & 72.42 (14.75) & 118.72 (10.32) & 0.19\\
    \end{tabular}
    \end{table}
\end{center}

\subsection{\RQOne}

\begin{figure}
\centering
\begin{subfigure}{0.49\textwidth}
        \includegraphics[width=\textwidth]{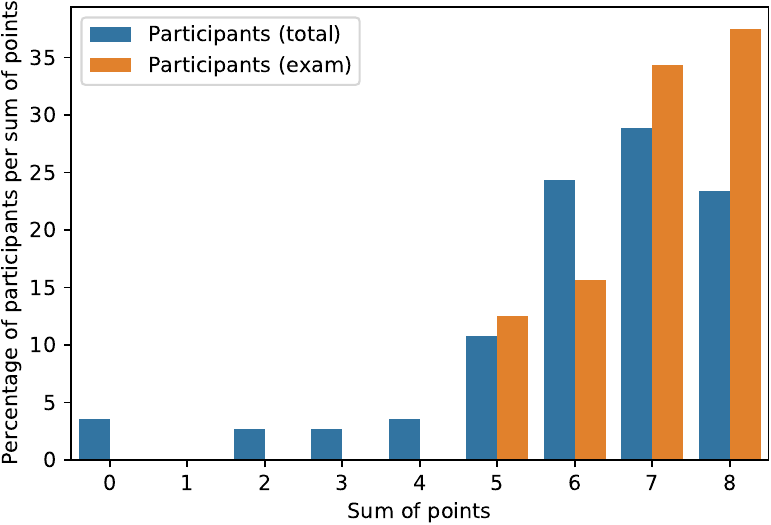}
          \caption{}
          \label{fig:barplot_score}
      \end{subfigure}
      \begin{subfigure}{0.49\textwidth}
        \includegraphics[width=\textwidth]{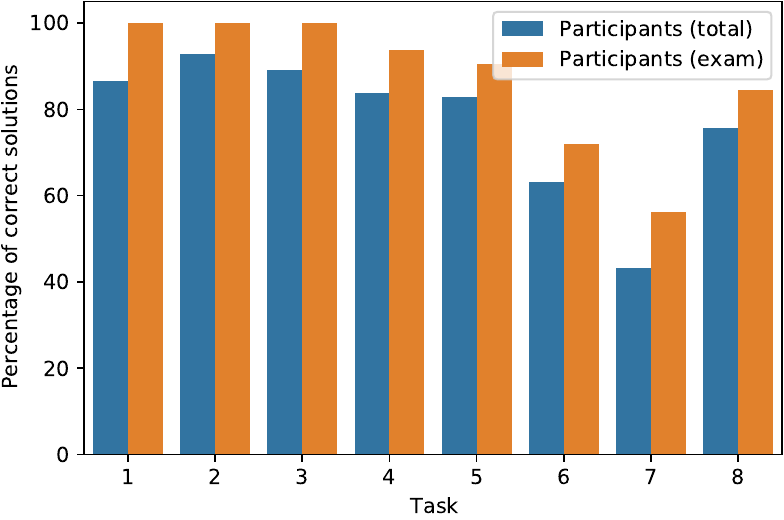}
          \caption{}
          \label{fig:barplot_per_task}
      \end{subfigure}
\caption{(a) sum of points for all participants and those who took part in the final exam. (b) number of correct solutions per task.}
\label{fig:histogram}
\end{figure}

The mean score for all 111 students who participated in the Early-MM test was 6.17 (sd = 1.86), with a maximum of 8 points. The majority of the students received at least 75\,\% of the points, indicating a ceiling effect. In Figure~\ref{fig:histogram}, we visualize the results. Considering the responses by task (Fig.\ \ref{fig:barplot_per_task}), the majority of the students is able to correctly solve them, with the second task being the easiest. Thus, most of the students in the course appear to have mastered the preoperational stage, but may not have fully reached the concrete operational stage. Interestingly, Tasks 6 and 7 (providing a natural-language description of source code) appeared to be most difficult, but not Task 8, which was intended to be the most difficult tasks (i.e., implement a swap).

Of these 111, only 32 students finished the programming course by taking the final exam of the DS course. The mean for these 32 is 6.97 points (sd = 1.03), so the ceiling effect is even more pronounced. Of these 32, all scored 5 or more points, with the majority scoring 7 or 8 points (23; cf.\ Fig.\ \ref{fig:barplot_score}). Regarding the points per task, the first 3 tasks have been answered correctly by all participants. Furthermore, the students are performing approximately 10\,\% better, compared to all participants who completed the Early-MM test.

%Anzahl Punkte pro Aufgabe der 32 Klausurteilnehmenden
% Aufgabe 7 -> schwierigste Aufgabe 
% 17/32 = 56,25% 
% 48/111 = 43,24%
%1	32
%2	32
%3	32
%4	30
%5	29
%6	23
%7	18
%8	27

%1. Wie viele Studierende wie viele Punkte haben (was schon da ist)
% 2. Wie verteilen sich die Punkte pro Aufgabe (wenn wir Zeit haben)

%To answer RQ1, we correlate the points in the test with the points in the final exam. 
The correlation with the exam is r = 0.23 (p = 0.204). Thus, the performance in the Early-MM test has a weak correlation with the performance in the final exam of the course.

\fbox{The Early-MM test can predict the success of the programming course to a small degree.}

% \begin{itemize}
%     \item CS1 (Marcs Zahlen)
%     \begin{itemize}
%         \item n = 114
%         \item EIOS test: mean = 6.24, sd = 1.8, median = 7
%         %\item Programming tasks: mean = 19.96, sd = 11.5, median = 24
%     \end{itemize}
%     \item CS2
%     \begin{itemize}
%         \item n = 44 (mit Abbrechern), n = 33 (ohne Abbrecher)
%     \end{itemize}
% \end{itemize}
% \begin{center}
%     \begin{tabular}{c c c c}
%         \hline
%          EIOS score & n (attended exam) & mean exam [in \%]& did not attend exam \\
%          \hline
%          < 5 & 2 & 68.00 & 2 \\
%          5 & 2 & 80.25 & 3 \\
%          6 & 4 & 70.00 & 2 \\
%          7 & 11 & 68.45 & 1 \\
%          8 & 14 & 81.96 & 3 \\
%     \end{tabular}
% \end{center} \\

%\begin{figure}
%\centering
%\begin{subfigure}{0.49\textwidth}
%    \includegraphics[width=\textwidth]{figures/barplot_p%er_task.pdf}
%          \caption{}
%          \label{fig:barplot_per_task_all}
%      \end{subfigure}
%      \begin{subfigure}{0.49\textwidth}
%        \includegraphics[width=\textwidth]{figures/barpl%ot_per_task_32.pdf}
%          \caption{}
%          \label{fig:barplot_per_task_exam}
%      \end{subfigure}
%\caption{Number of Correct Answers per Task - %\ref{fig:barplot_per_task_all} - All 111 Participants, %\ref{fig:barplot_per_task_exam} - 32 Exam Participants}
%\label{fig:barplot_per_task}
%\end{figure}

\subsection{\RQTwo}
To answer RQ2, we used three different tests: The C-Test, the R-Voc and the P-Voc (cf.\ Section~\ref{sec:material}). 

50 students completed the C-Test. The mean score was 76.90 points (a maximum of 100 points could be reached), with a standard deviation of 13.89. The median was 81. From these 50 participants, 36 took the final exam. The mean for these 36 is 80.11 points (sd = 12.03) and the median is 84. Most students (31 of 50) are in the C1 level, which indicates a ceiling effect.

The R-Voc was completed by 39 students. We removed the data of one student, because they only received 24 points (from a maximum of 150 points), which is more than two standard deviations below the mean. The remaining 38 students received on average 129.76 points (sd = 18.09), and the median was 134.5 points. 27 students took part in the final exam. For these 27, the mean was 130.74 points (sd = 14.41) and the median was 134. Most students (28 of 38) are in the B2 level, indicating a ceiling effect.

The P-Voc had 35 participants. Out of 90 points, the mean was 61.77 points (sd = 11.45) and the median was 62 points. Out of the 35 participants, 26 completed the exam. They have a mean of 62.38 points (sd = 11.85) and the median was 61 points. The ceiling effect here is less pronounced: While on average, the participants still reached far more than half the points, only 7 students passed the highest rank and are at the B2 level.

%For the C-Test and the R-Voc, we see a clear ceiling effect. The results of the C-Test place over half of the participants in the CEFR level C1, which is the highest level it can assess~\cite{}. The situation is similar with the R-Voc. When taking a look at the participants' points per level, the mean was over 24 points (which corresponds to 80\,\% of the points per level) for all levels except Level 3, where it is only slightly lower (23.74). This means that, on average, the participants passed all levels of the P-Voc, placing them at a B2 reading level in the CEFR, which is the highest that the test can assess.
%For the P-Voc, while the participants on average still reached 68\,\% of the possible points, when looking at the individual levels, only Level 1 and 2 have on average over 80\,\% of the points, which is 14. 
%such that students show a performance well above the mean. This is not surprising, given that our sample consists of students, who typically have higher cognitive skills than the average human population. 

Noteworthy is also that, when considering only the students who completed the final exam, the test scores are a bit higher.

To answer RQ2, we correlate the points of each of the three language-skill tests with the points in the final exam. For all tests, the correlations are in the weak range, starting with 0.20 for the C-Test (p = 0.251), 0.28 for R-Voc (p = 0.159), and 0.27 for P-Voc (p = 0.186).

\fbox{Language skills can predict the success of the programming course to a small degree.}

\subsection{\RQThree}

49 students participated in the d2-R. For our participants, the mean was 57.12 (sd = 8.99). 36 finished the DS course by taking the final exam. Of these 36, the mean was 57.11 (sd = 9.27). As with the language skill tests, the students performed above average (the standardized mean is 50 points, sd = 10), although the higher performance was not so pronounced as for the language-skill tests or the Early-MM test.

The correlation between the d2-R and the exam is 0.09 (p = 0.584). 

\fbox{Selected and sustained attention cannot predict the success of the programming course.}

\subsection{\RQFour}

We received the response of 35 students for the DESIGMA-A. One participant took the test twice, and we only kept the response of the first attempt. The mean was 115.97 (sd = 12.51). 25 students finished the DS course by taking the final exam. Of these 25, the mean was 118.72 (sd = 10.32).

As for the language tests and d2 tests, the students performed above average, even one standard deviation above, showing their higher than normal intelligence as measured with this test. As for the Early-MM and the language skills, the performance of the students who finished the DS course is a bit higher than for all students who completed the DESIGMA-A.

The correlation between the DESIGMA-A and the exam is 0.19 (p = 0.373).

\fbox{Fluid intelligence can predict the success of the programming course to a small degree.}

\subsection{\RQFive}

\begin{center}
\begin{table}

    \caption{Correlation matrix of all tests. **\begin{math}p < .01\end{math}, *\begin{math}p < .05\end{math}. Exact p-values provided on the project's Web site.}
    \label{tab:RQ5}
    \begin{tabular}{l | c c c c c c c}
    \toprule
         & d2-R & DESIGMA-A & C-Test & R-Voc & P-Voc & Early-MM & Exam\\
        \midrule
        %\hline
        d2-R & 1.00 & 0.12 & -0.16 & 0.20 & -0.14 & -0.14 & 0.29 \\

        DESIGMA-A & 0.12 & 1.00 & -0.17 & 0.41 & 0.20 & 0.36 & 0.23 \\

        C-Test & -0.16 & -0.17 & 1.00 & 0.52* & 0.83** & -0.05 & 0.13 \\

        R-Voc & 0.20 & 0.41 & 0.52* & 1.00 & 0.68** & -0.06 & 0.24 \\

        P-Voc & -0.14 & 0.20 & 0.83** & 0.68** & 1.00 & -0.11 & 0.24 \\
        
        Early-MM & -0.14 & 0.36 & -0.05 & -0.06 & -0.11 & 1.00 & 0.28 \\

        Exam & 0.29 & 0.23 & 0.13 & 0.24 & 0.24 & 0.28 & 1.00
        %\bottomrule
    \end{tabular}
    \end{table}
\end{center}

To answer RQ5, we calculated the correlations for all tests from RQ1 to RQ4. The plan was to integrate the results into a stepwise regression model. However, only 18 students completed all tests and the final exam, which is not sufficient for computing a stepwise regression model. Instead, we take a detailed look at each correlation one by one. In Table~\ref{tab:RQ5}, we present the correlation matrix of all tests and the final exam. The mean score for the 18 students in the exam was 73.03 (sd = 16.63), which is higher than the mean for all students (66.61).

Early-MM has no considerable correlations with the other tests, except DESIGMA-A, which means the uniqueness it provides as predictor appears rather high. The correlation with the exam is low (0.28), but still among the highest when compared to the other predictors. Thus, even though there is a ceiling effect, the Early-MM might be a good candidate for the early warning system.

The language tests correlate highly with each other (\begin{math}r >= 0.52\end{math}), especially C-Test and P-Voc, suggesting that these two tests do not add much uniqueness as predictors. Thus, one of these tests would suffice for the early warning system. Of these two, the P-Voc has a higher correlation with the exam, but the C-Test has a lower correlation with the DESIGMA-A. Since both tests take about the same amount of time to apply, either of these tests might be a good candidate for the early warning system.

The d2-R only has low correlations with the other tests, so from that perspective, is a good candidate for the early warning system. Interestingly, taking only the students into account who completed all the tests as well as the final exam, the correlations changes drastically, from almost zero to an (cf.\ Table~\ref{tab:summary}) almost medium correlation. Thus, it may not be a good predictor for all students of a course, but only for the ones who are highly motivated. Hence, the d2-R may not be suitable for our early warning system, as it cannot differentiate among the weaker students, which we want to identify with it.
%In other words, the d2 test might only be able to differentiate among the highly motivated students.
% Idee: Kann bei der Allgemeinheit unserer Population nicht mehr gut differenzieren kann, aber vielleicht bei den Leuten, die eh schon gut sind; passt dann ggf nicht so gut zum Frühwarnsystem, weil eh nur bei denen differenziert wird, die eh keine Probleme haben

The DESIGMA-A has a medium correlation with Early-MM, but a low correlation with the exam. Additionally, it has a medium correlation with R-Voc. Low correlations with the other tests also make it a good predictor regarding uniqueness. In contrast to Early-MM, it can be applied before a programming course starts, since it does not require any prior training. 
% Thus, it can be helpful in identifying students who might need more support to learn programming and maybe assign them more specific training tasks. However, it takes a rather long time to complete the test (almost an hour) and thus puts considerable strain on the students. 
% (though it is notably bigger with this sample compared to RQ4) (wobei das vermutlich eher ein Punkt für die Diskussion ist(?))

\noindent\fbox{\begin{minipage}{\dimexpr\textwidth-2\fboxsep-2\fboxrule\relax}
\raggedright %andere Option: \center
Possible candidate tests for an early warning system are the Early-MM and/or the DESIGMA-A, plus one language-skill test, such as the C-Test or the P-Voc.
\end{minipage}}

%\fbox{Possible candidate tests for an early warning system are a %the Early-MM and/or the DESIGMA-A, plus one language-skill test, %such as the C-Test or the P-Voc.}
% Können wir leider nicht machen, da zu wenige Daten. Geplant war ein stepwise regression model. Stattdessen schauen wir uns jetzt die einzelnen Korrelation an.
% Ok, da produktiv und c test hoch korrelieren, trägt jeder einzelne nicht viel uniqueness bei. es reicht, wenn wir einen ins frühwarnsystem mit aufnehmen. Schauen wir uns die korrelation mit klausur an, scheint productive eine bessere vorhersage-kraft zu haben, weslhab dieser am bestne ins frühwarnsystem passen könnte.

% p values ergänzen (der Vollständigkeit halber, betrachten wir aber nicht), beschreiben was man in der tabelle sieht
% desigma x R-Voc p = 0.09

\subsection{\RQSix}

47 students participated in all parts of the syntax drill-and-practice exercises. One participant submitted partially blank responses, so we assume his responses are not genuine, so we remove the data as an outlier. Thus, our sample consists of 46 students. Overall, the students received an average score of 95.67\,\%. Students' performance improved from exercise to exercise, that is, from 92.16\,\% to 98.17\,\%. 

%Per error category, there is a big difference in general syntactic errors that are independent of the programming constructs, such as wrong or missing brackets. From the first test, conditions, to the last test, while loop, these have drastically decreased. However, there is not much difference in errors that are specific to a programming construct, such as assignment operator in conditions, incrementing a variable, or terminating the condition with a semicolon in a do-while loop. 

\begin{figure}
\centering
\begin{subfigure}{0.49\textwidth}
        \includegraphics[width=\textwidth]{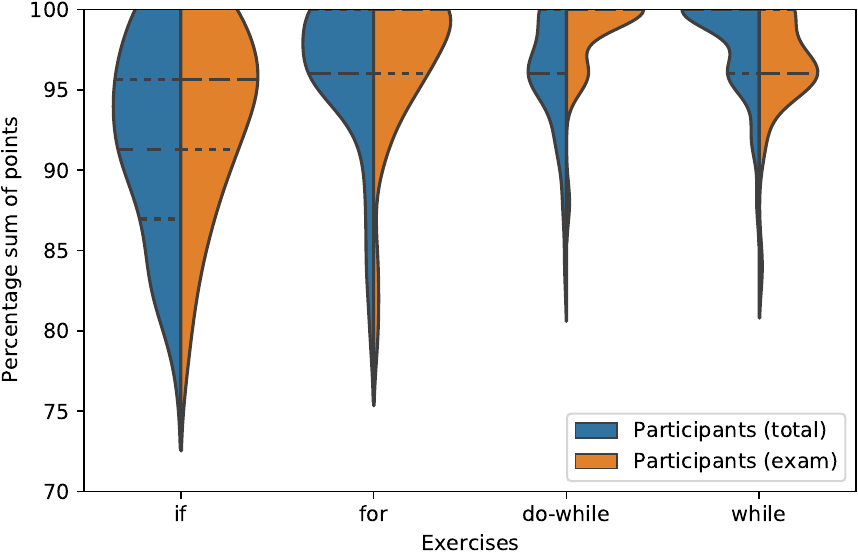}
          \caption{}
          \label{fig:grouped_violinplot}
      \end{subfigure}
      \begin{subfigure}{0.49\textwidth}
        \includegraphics[width=\textwidth]{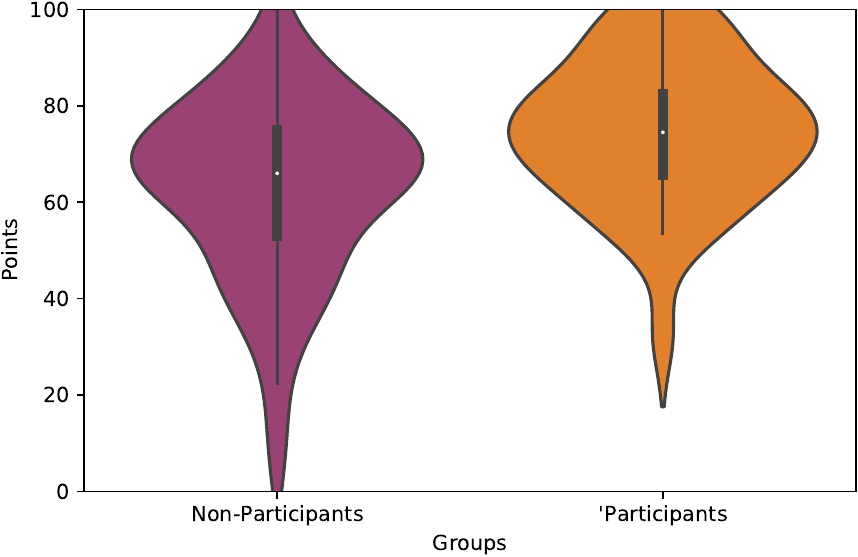}
          \caption{}
          \label{fig:exam_violinplot}
      \end{subfigure}
\caption{(a) Percentage of correct submissions per exercise type. (b) Points in the final exam, divided into students who participated in the syntax drill-and-practice exercises (orange) and students who did not participate (purple).}
\label{fig:violinplot}
\end{figure}

Of these 46, 33 students participated in the final exam. The average score in the syntax drill-and-practice exercises of these participants was 96.47\,\% (sd = 2.69), so comparable to the entire sample (95.67\,\%, sd = 2.89). 
Broken down per programming construct, the number of points also increases per construct (93.41\,\% to 98.79\,\%), with a still stand from for loops (96.97\,\%) to do-while loops (96.73\,\%).
% achieved increases almost continuously from task part to task part. The average result for the conditionals was 93.41\,\% (range: 78.26\,\% - 100\,\%). After that, the result increased for for-loops (mean = 96.97\,\%, range: 80\,\% - 100\,\%).  A slight decrease took place for do-while-loops, with an average result of 96.73\,\% (range: 84\,\% - 100\,\%). Finally, there was again an increase in while-loops, which was somewhat higher, with an average percentage of  98.79\,\% (range: 88\,\% - 100\,\%) of the points achieved.

% In the first task, participants made an average of 1.48 errors. In contrast, there were only 0.30 errors in the last test. For the wile loops, there was a short increase to 0.82 errors. Overall, there was a substantial decrease in errors from the first to the last part of the syntax drill-and-practice exercises. The distribution of the different error categories was not equally distributed per task part.

Regarding error category, there are some interesting observations: First, there were many mistakes students made with errors of the bracket category at the first exercise (33), which drastically was reduced for the last exercise (2). Second, for errors in the loop counter category, the only mistakes students made was for the for loops, which was the first construct in which these errors occurred. Last, for the errors that cannot be assigned to a specific category, in the last exercises (while loop), students made most of the mistakes (8 of 10). This specific error was that the loop counter (a.k.a., stepper) was missing, making the loop an infinite loop. Thus, this may actually not be a syntactical error, but more a semantic error, or a combination of both. 

%Due to a different distribution of error categories between the different tasks, a comparison was not possible
%Overall, the number of errors has been significantly reduced to one fifth of the errors in the first task (if: 49 mistakes, for: 25 mistakes, do-while: 27 mistakes, while: 10 mistakes) 
%After a lot of mistakes were made at the beginning of the bracket category, not even 10% of the mistakes were made at the end compared to the first task (if: 33 mistakes / while: 2 mistakes)
%In the loop counter category, only errors were made in the for-loop test
%In the while loops, errors were almost only made in the Other Category. A reason for this is given in the discussion (8 of 10 mistakes)

To evaluate whether the syntax drill-and-practice exercises have an effect on the acquired programming skills, we compare the points in the exam for students who participated in the syntax drill-and-practice exercises with students who did not participate. Students who completed the syntax drill-and-practice exercises received on average 74.61 points (sd = 15.04; the maximum is 100 points). By contrast, the 69 students who did not participate in the syntax drill-and-practice exercises scored an average of 62.96 points (sd = 18.5) in the exam, so considerably lower. This difference is significant (Mann-Whitney-U test; U = 1555.5, p = 0.003) and indicates a medium effect size of 0.37 (Cliff's Delta).
% The data does not follow a normal distribution (Shapiro-Wilk, W = 0.97, p = 0.488). A Mann-Whitney-U test revealed that the difference is significant (U =  1555.5, p = 0.003), with a medium effect size of 0.37 (Cliff's Delta).

\fbox{Syntax drill-and-practice exercises can help students improve their programming skill.}

%Mittelwerte, Standardabweichung, %Violinplots/boxplots/histogramm
%Marc: Programmieraufgaben können nicht gut differenzieren -> sind nicht gut als Frühwarnsystem, eher die kognitiven Tests

% \subsection{Inferenzstatistik}
% %Shapiro-Wilk-Test
% %Syntax-Teilnehmer: p = 0.44 / Nichtteilnehmer: p = 0.01
% %Keine Normalverteilung = MWU 
% % MWU =  U-val: 1617.5 p: 0.001
% Korrelations, Analyse der Ko-Korrelation

% Wer hat schlecht in CTest und d2 abgeschnitten? -> unteren 20\%
% Wer von denen hat Syntax-Training gemacht? Hat das geholfen?
%starker ceiling effekt, kein zusammenhang mit exam

\section{Discussion} \label{discussion}

% \begin{itemize}
%     \item moderate correlation of exam with d2-R: seems to fit previous findings regarding working memory/attention
%     \item low correlation between exam and DESIGMA-A: does not fit previous studies who generally find high correlation with fluid intelligence
%     \item ceiling effect for language tests. Interesting: The test with the smallesst ceiling effect has the highest correlation with the exam
%     \item generally moderate correlations with language tests: seems to fit previous findings
% \end{itemize}
% %Klausur + Syntaxaufgaben - interessant, dass 14 von 48 nicht mitgeschrieben haben 
% %OPAL DS sind 172 eingetragen 
% %Syntaxaufgaben - primäres Ziel Reduzierung der Cognitive Load
% Mögliche Prädiktoren wären: Ein Wortschatztest, d2, evtl. Desigma, aber so viel Aufwand, diesen durchzuführen,
% Diese einfachen Programmieraufgaben können nicht Erfolg vorhersagen vermutlich, weil hier jetzt Informatikstudierende dabei sind
% Funkioniert unser Frühwarnsystem?Jein;
%     Marcs Aufgaben: nein, die sind zu leicht für unsere Population -> aktuelle Studien mit schwereren Aufgaben
%     Kognitive Fähigkeiten? -> Wortschatz und d2 scheint gut zu funktionieren
%     Helfen Syntaxaufgaben? -> Vielleicht

\subsection{\RQOne}

We found that students' ability to develop a suitable mental model at the beginning of a programming course can predict the performance of acquired programming skill to a certain degree. Given that it is also easy to apply, it is a suitable test for our early warning system. In contrast to the results by Ahadi and others, we found that most students had reached the concrete operational state and could implement a swap of values between two variables. One reason might be that many of our students had some previous programming experience: On average, our students have been programming for 1.8 years, while Ahadi and others assume that their students do not have any programming experience~\cite{ahadi_falling_2014}. Thus, the students of our sample had considerably more time to develop a suitable mental model of programming. This might also explain why we observe a lower correlation, that is, in our sample, the Early-MM test cannot differentiate sufficiently among our more experienced students. In the second instance of the programming course, we have extended the task to cover more advanced aspects of programming, such as logical expressions and arrays. Data collection is currently ongoing, so we cannot give an impression of how the revised version of the Early-MM works.

Interestingly, we observed that two of the questions are especially difficult for our students. In these questions, students had to give a natural-language description of what a piece of source code was doing. Especially describing the swap proved more difficult than implementing one. This could hint to a problem that students have difficulties in comprehending code, rather than producing it. Another explanation could be that, since the swap is such a basic algorithm, students might be used to implementing it without actually having a natural-language representation. This result was also observed by Ahadi and others for one of their sample of students, who also scored better at implementing the swap than describing it, although the difference was less pronounced (i.e., 50\,\%, 44\,\%, and 53\,\% for Tasks 6, 7, and 8, respectively; in the other sample, the implementation of the swap was the most difficult, i.e., 42\,\%, 33\,\%, and 31\,\% for Tasks 6, 7, and 8, respectively). Diving deeper, we found that common mistakes were that participants (i) assumed that all variables would get the same values, (ii) named the temporary variable to be swapped, instead of the target variables, or (iii) described the swap on a too high abstraction level. Thus, it might be helpful for students to focus more attention on the purpose of code, so to also include program-comprehension task in programming courses~\cite{izu_fostering_2019}. Although students were able to read source code and trace variable values, they could not describe the exact purpose of the code, so failed to integrate the statements into semantic chunks, an important part of bottom-up program comprehension~\cite{Pennington87}. Hence, students have not yet reached the concrete operational stage and their mental model is not sufficiently developed to understand more complex algorithms~\cite{lister_cognitive_2020}.

\subsection{\RQTwo}

All the language tests showed a certain relationship with the performance in the final exam. Thus, we could strengthen the result of previous studies, even though the strength of the relationship is weaker in our data than, for example, compared to Prat and others~\cite{prat2020relating}. One reason could be that most participants received an above average score, so the language tests were too easy for our intended population of computer science majors. Additionally, Prat and others used a different operationalization of language skills that predicts whether someone is able to learn a new language, whereas we tested whether someone already is proficient in a foreign language. 

% Tests that can differentiate better among a higher skilled population appear to be better predictors for acquiring programming skills. Thus, i
In the future, we intend to look for tests that are more difficult, but still easy to apply. Currently, there is a C-Test being constructed that is more difficult, and we plan to integrate this test in our ongoing data collection among students of the two-semester programming course.

The advantage of our tests is that they are very quick to apply and easy to access. That is a problem for the test used by Prat and others~\cite{prat2020relating}, which is not available for research, so it is not useful for our early warning system.

\subsection{\RQThree}

It is unclear how selective and sustained attention is related to programming skill. Considering the relationship between students who completed the d2-R test and the exam, the correlation is not different from zero, but considering the correlation of students who completed all tests and the exam, the correlation increases and could almost be considered as medium (0.29, cf.\ Table~\ref{tab:summary}). One possible interpretation could be that the d2-R test can only differentiate among the students who are highly motivated to perform well in a course and take part in all learning opportunities. However, since this test cannot differentiate between the students that we want to identify with it, it is not suitable for our early warning system. This is in line with the interpretation by Ambrosio and others, who concluded that simple attention tasks might be too easy to really differentiate between programming students~\cite{ambrosio_identifying_2011}. Even though they used a different operationalization of attention, we observed a similar result.

% Ambrosio et al.:
% \begin{itemize}
%     \item Goal: "One of its primary purposes it to identify the skills associated to programming, ergo, central in CT"
%     \item Result: r = .273(exam) bzw. .310(final grade). "this possibly means that [simple attention or calculus tasks], being very basic skills, with no inherent significant difficulty, all students attain a very similar level of performance, and therefore, they are not relevant  to the differentiating goal of our study"
%     \item Conclusion: "to confirm the lack of correlation between programming and attention to details, a different perception test should replace the Tool Matching test, which may not have shown correlation due to the test format and lack of face validity"
% \end{itemize}

% \begin{itemize}
%     \item Correlation fluctuates a lot when changing the sample (vgl. Results von RQ3 und RQ5)
%     \item Ambrosio et al also found that attention to detail only had a low correlation with their students' final exam (r = 0.273). They used a very different test, but our results seem to be more or less in line with theirs.
% \end{itemize}

\subsection{\RQFour}

We found weak correlations between the fluid intelligence and programming skill, strengthening previous findings that higher fluid intelligence is useful when acquiring programming skills. As with the other cognitive skills, this is less pronounced compared to other studies, which often reported medium to high correlations between fluid intelligence and programming skill~\cite{ambrosio_identifying_2011, prat2020relating, roman2017cognitive}. The disparity between these and our results might be due to the difference in operationalization of fluid intelligence, since the DESIGMA-A uses a solution-composition approach to figural matrices (cf. Section~\ref{sec:material}). The operationalization of programming skill also differs between studies, so it is difficult to truly compare the results.

%For example, Ambrosio and others found medium correlations between general intelligence and programming skill, as well as medium to high correlations between spatial reasoning and programming skill~\cite{ambrosio_identifying_2011}.Endres and other similarly report high correlations between spatial ability and programming skill~\cite{endres2021read}.Rom\'{a}n-Gonz\'{a}lez and others found medium correlations when correlating Computational Thinking with spatial and reasoning ability, as well as high correlation with problem-solving~\cite{roman2017cognitive}. The disparity between these and our results might be due to the difference in operationalization, as spatial reasoning is different from figural inductive reasoning. However, Prat and others also found high correlations between programming learning outcomes and fluid intelligence, using a figural inductive reasoning test specifically~\cite{prat2020relating}. It might then be that the test environment or the special distractor-less method of the DESIGMA-A play a role.

Additionally, we found that fluid intelligence and the ability to build a correct early mental model of programming have a medium correlation. Thus, in the beginning of a programming course, a higher fluid intelligence appears to be helpful, but this advantage may vanish during the programming course. This is also in line with neuro-imaging studies, in which researchers observed that experience in programming leads to a more specific brain activation during programming tasks, compared to novices, who typically show a pattern of individual cognitive processes and less efficient use of neuronal resources~\cite{helmlinger2020programming,siegmund_measuring_2017,Crk:2015:Brains}. Thus, the DESIGMA-A might be a good predictor for how students start a programming course, but might be less relevant when students have acquired a certain skill level in programming.
% While plausible, this contradicts the findings of Helmlinger and others, who find that participants with programming experience had higher fluid intelligence than non-programmers~\cite{helmlinger2020programming}. It would be interesting to see if our early warning system would be different when ... 

Hence, the DESIGMA-A could be a useful predictor for our case, but is rather tedious to apply, as it takes over one hour to administer. Recently, a short version of the DESIGMA has been released, which only takes up to 23 minutes~\cite{desigma-s}, so less than half the time of the long version. With this reduced effort, the short version of DESIGMA may be of help for us, so we have integrated this version in our ongoing study.

\subsection{\RQFive}

Unfortunately, the small sample size for students who participated in all tests and the final exam did not allow us to compute a statistical model, such as a stepwise regression model, to build an early warning system. Nevertheless, we could get a first impression of possible candidates by manually analyzing the correlations among all tests and exam scores.

Especially language skills and a correct early mental model appear to be able to predict whether students have developed programming skills at the end of a two-semester programming course. In addition, fluid intelligence may also be a suitable predictor. The advantage of assessing cognitive skills, rather than programming skills, is that this can happen directly at the beginning of a programming course or even before that, so that students who have a higher risk of failing a course can be made aware of that risk early on. This way, they know that they should be careful not to fall behind, because it is difficult to catch up. 
Furthermore, these students can receive special course offers, for example, in terms of tailored assignments or especially trained teaching assistants, which can help students not to fall behind.

In a sample where students have some previous exposure to programming, the Early-MM test can also be applied directly at the beginning or before the actual course, with the same goal of identifying students who are at high risk of struggling. When applied a few weeks into a programming course, it may need some adjustments to fit a more experienced demographic. With the integration of programming education in more and more middle and high schools, it is highly likely that more and more students start a major and already have some programming skills, so an adjusted test is definitely useful in the future.

% \begin{itemize}
%     \item Fazit: Early-MM und lanugage tests sind vielversprechend, aber wir brauchen bessere tests (für eine demographie mit höheren kognitiven Fähigkeiten). Allerdings wäre es natürlich auch interessant zu sehen, wie die Situation ist, wenn man die Tests am Anfang von AP macht, statt in DS.
%     \item Die Tests messen jeweils unterschiedliche Sachen, abgesehen von den language tests, bei denen man sich wohl mindestens einen sparen könnte, wenn nicht sogar zwei. Hängt jetzt wohl davon ab, welchen wir in einer schwierigeren Ausführung bekommen können.
%     \item Die Wortschatztests korrelieren besser mit DESIMGA-A, aber warum?
%     \item DESIGMA-A ist vor allem wegen seiner Länge nicht so schön, aber es gibt eine kürzere Version
%     \item wenn die einzelnen skills etwas predicten können, kann das auch die Kombination
%     \item in Zukunft wärs natürlich wichtig, dann auf die ergebnisse des frühwarnsystems auch differenziert reagieren zu können. Weitere Studien vonnöten.
% \end{itemize}

\subsection{\RQSix}
%if: Other Category Wrong if-else if-else
%do-while: : instead of ;
%While: Other Category - missing Counter

We found that syntax drill-and-practice exercises can improve students' programming skill at the end of a two-year programming course. Unfortunately, our data does not allow us to draw conclusions on whether these exercises are especially useful for students who have a low score in the early warning system, as we only have 12 students who participated in all tests, the exam, and the syntax drill-and-practice exercises. Nevertheless, since these exercises do not require much effort, neither for students nor for educators, and since there is a measurable positive effect, we are currently integrating these exercises into the course material. Additionally, we are increasing their difficulty level, so that not only the inexperienced students profit from them, but that also more skilled students can improve their programming skills.

Interestingly, we found that, for later syntax drill-and-practice exercises, the number of mistakes decreases across error categories, especially for bracket errors (33 vs.\ 2) and loop counter errors (only for for loops). This could mean that, when students have familiarized themselves with certain errors, they can transfer that to different programming constructs. Thus, it may be possible to create a more efficient version of syntax drill-and-practice exercises, such that programming constructs and error categories can be combined and the number of items can be decreased.

More efficient syntax drill-and-practice exercises also means that we can include more programming exercises that build on the syntax drill-and-practice exercises, such as fill-in-the-blank tasks (e.g., used by Kramer and others~\cite{kramer_identifying_2019} and Kyaw and others~\cite{kyaw_study_2021}) or parsons puzzle~\cite{denny_evaluation_2008, ericson_solving_2017, parsons_parsons_2006}. Additionally, errors that can be a combination of both, syntactical and semantic errors, can be integrated, such as the missing loop counter that we accidentally as syntax error, but might actually turn out to be a semantic error, or a combination of both.

\section{Implications} \label{implications}

The results of our study are promising for an early warning system. It can be applied even before actual programming courses start to help students by assigning them to specially trained tutors or provide them with tailored training tasks. These could be similar to the tasks defined by Thurner and others, who provide tailored tasks based on performance of preceding programming tasks~\cite{thurner2017concept}. However, this can only happen a few weeks into a course, but with an early warning system, this process can be started even before that, so that we do not lose valuable time at the beginning of a course.

% We need tests that can better differentiate among the high skills. That might have been also an issue in the other studies, ...

Syntax exercises help, even when they are applied later in the course. Of course, earlier might even be more helpful, as found by the use of Phanon by Edwards and others in several studies~\cite{edwards_separation_2018, edwards_syntax_2020, sullivan_student_2021, ly_revisiting_2021}. Additionally, special training in reading and spatial skills can also help improve programming skill, and these can happen even before a programming course~\cite{endres2021read}.

Despite having rather preliminary data, we are hopeful that our candidate early warning system is a good starting point to identify struggling students so early that we can still implement measures to support them in overcoming their struggles. Noteworthy are the consistent ceiling effects for all tests, meaning that, for all tests, our sample performed well above the mean. This can be expected, considering that our sample typically has higher cognitive skills than the average population (i.e., including persons not pursuing a degree in higher education). For building an early warning system, it may increase predictive power if we include tests for cognitive skills that are designed for a population of higher cognitive skills and can differentiate better among students. This is what we currently do in our ongoing studies.

Another interesting trend is that students who persevered in the course until the final exam tend to score better in all the tests that we applied. Additionally, students who took part in the study for at least one test or even all of the tests have a higher mean in the final exam. This may be because these students actually have higher cognitive skills or acquired better programming skills, but might also hint at a selection bias, such that only the highly motivated students participated (participation in the tests was voluntary). There might also be a moderator effect, such that higher cognitive skills lead to higher motivation in the course, because students have an easier time to understand the course material. In future studies, we hope that we can nudge more students to complete these tests, so that we can differentiate the true effect of cognitive skills and motivation.

To conclude, even though the early warning system does not solve the difficulties of teaching programming, it is still an important piece in the puzzle to do so, because when detected early, it is possible to provide support for struggling students to overcome their struggles and successfully acquire programming skills.

% With cognitive tests that can better differentiate among the high-skilled population of students, the early warning system might become a useful tool in offering tailored support, so that more students can develop useful programming skill.

% In future work, we want to refine our testing methods, especially through the usage of tests that can better differentiate among the target demographic.
%We could also consider taking other factors than just cognitive skills into account, such as personality, as there are hints that different personality types might have different strengths and weaknesses when it comes to programming learning
%evlt. personality aufgreifen

% Implication: All RQs zusammen im Big Picture diskutieren.
% \begin{itemize}
%     \item Testing different aptitudes outside of programming, like language skills, seems promising, but test that better suit our demographic are needed (though it would be interesting to test other demographics)
%     \item Syntax exercises seem to be promising in helping students increase their programming skill
%     \item currently we cannot infer that syntax exercises help those who do less well in language tests due to a lack of data - the combination of identifying struggling students and applying suitable treatment needs to be further explored
% \end{itemize}

% \section{Conclusion} \label{conclusion}

% \section{Acknowledgments}
\begin{acks}
  Many thanks to our students who participated in our studies.
\end{acks}

\bibliographystyle{ACM-Reference-Format}
\bibliography{references}

% %%
% %% If your work has an appendix, this is the place to put it.
% \appendix









\end{document}